\documentclass[runningheads]{llncs}
\usepackage[T1]{fontenc}
\usepackage{graphicx}
\begin{document}
%
\title{Toward Low-Latency Services over PON using OCDMA Private Networks}
%
%
\author{Steevy Joyce Cordette\inst{1}\orcidID{0000-0001-9769-1001} 
}
\authorrunning{S. J. Cordette}
%
\institute{Senior Member, IEEE \\
\email{steevy.cordette@ieee.org}\\
}
\maketitle              
\begin{abstract}
An low-latency service scheme is proposed over Passive Optical Network (PON). The Optical Code Division Multiplexing Access (OCDMA) technique is used to define multiple private networks serving as Virtual GE-PON that mimic the service-based VLAN (S-VLAN) in the optical domain.

\keywords{Optical Communications   \and PON  \and Network  \and Low Latency.  \and CDMA}

\end{abstract}
\section{Introduction}

As emerging time-critical applications (e.g. autonomous driving, industrial automation, remote robotic surgery, 6G optical fronthaul, etc.), the passive optical network (PON) architecture faces growing challenges in ultra-reliable low-latency communication (URLLC) services (<1ms)\cite{ref_article1,ref_article3,ref_article4}. In recent years, novel schemes for low latency PON have been developed, notably frame-based dense burst allocation (FDBA), dividing by 4 the latency related to Dynamic Bandwidth Allocation (DBA)\cite{ref_article2}.Complementarily, to improve latency and reliability in Gigabit Ethernet PON (GE-PON), Virtual LANs (VLANs) can be employed to prioritize traffic for time-critical services, potentially at the expense of non-time-critical traffic. 


In this paper,  we introduce the concept of Virtual GE-PON to further minimize the latency of access network's for time-critical services, and propose a low latency strategy based on this architecture. 
Those Virtual GE-PON are based on multiple private networks using Optical Code Multiplexing Access (OCDMA).
The use of this service-oriented approach of OCDMA expands the concept of multiple private networks (PN) reported in \cite{ref_Gh_1}, where PNs of ONUs were demonstrated over the standard EPON.

\section{Network Architecture}
\begin{figure}
\includegraphics[width=\textwidth]{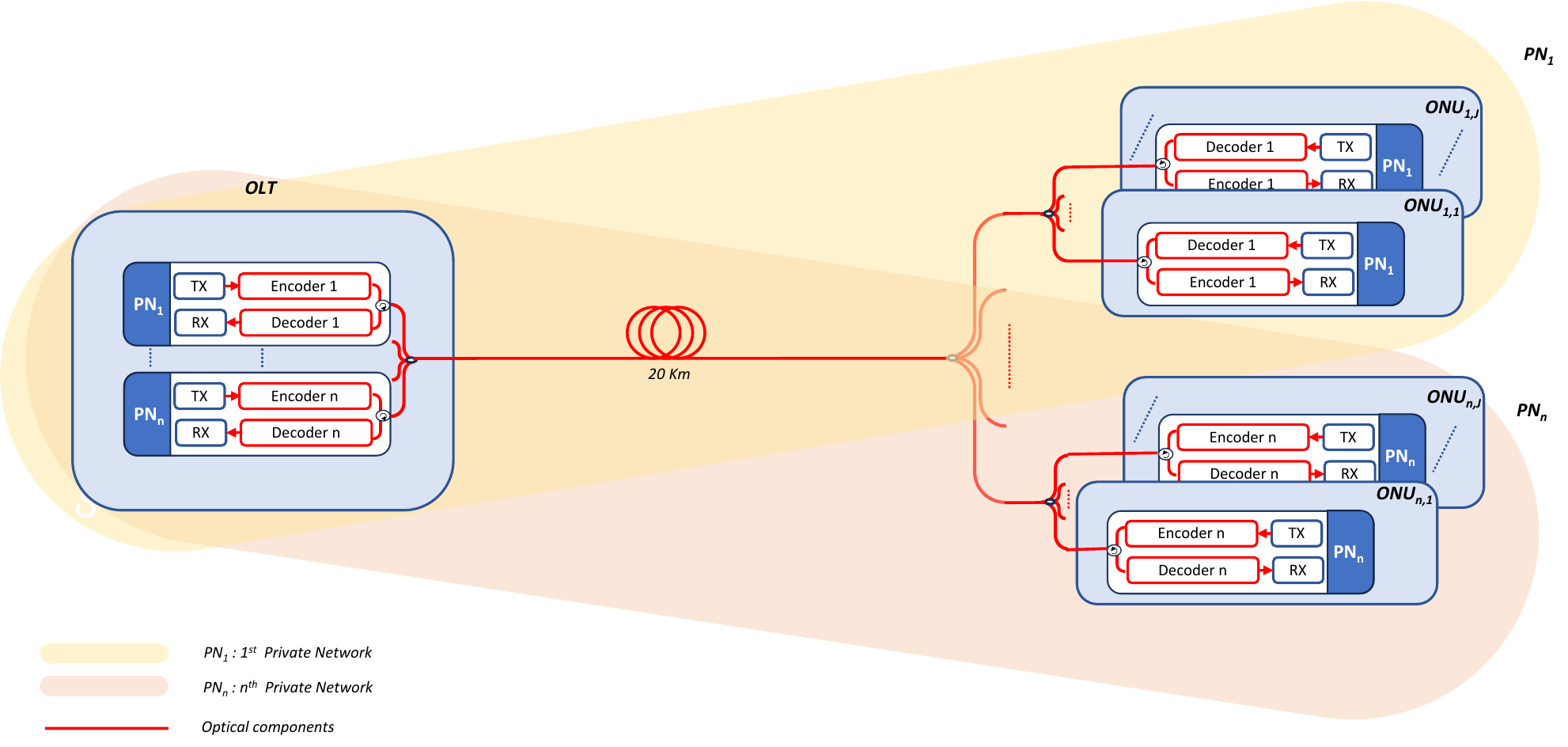}
\caption{Architecture of multiple optical private networks. Note: PN don't require to be on the same splitter branches; ONUs could be on Multiple PNs} \label{fig1}
\end{figure}

Fig.~\ref{fig1} shows the network architecture.  Unlike \cite{ref_Gh_1}\cite{ref_Gh_2}, the multiple private networks are chosen to include the Optical Line Terminal (OLT) (i.e., no Fiber Bragg Gating (FBG) reflector between the feeder fiber and the optical splitter). OLT and Optical Network Units (ONUs) are used to exchange upstream / downstream data at different wavelengths (1260\,nm and 1490\,nm, respectively).  
Additionally, the multiple optical PNs over the PON is achieved to operate at the wavelengths mentioned above. The OCDMA technique is used to implement the PNs by encoding/decoding each bit of a PN's data traffic with a codeword (corresponding to a given sequence of pulses) \cite{ref_Ihs}.

As a result,  a different codeword is assigned to each PN,  and used by the network element of the PN (OLT and ONUs) to establish a communication only detectable over the PN (based on the orthogonality performances of the set of code). 
From the experimental demonstration reported in \cite{ref_Gh_1,ref_Gh_2} of OCDMA-based multiple PN, and Direct-Sequence OCDMA (DS-OCDMA)\cite{ref_ware1,ref_ware2}, 2\,PNs could be implemented (with a penalty of 4 to 5\,dB at 10e-9 Bit Error Rate (BER)), and, respectively, a penalty of 8 to 10\,dB for 3\,PNs. For higher numbers of PN (i.e.,\,>3), electrical or optical thresholders can be used to eliminate MAI to achieve better BER performance \cite{ref_ware2,ref_NLT_Ihs_1,ref_NLT_1,ref_NLT_2,ref_NLT_3}.

\section{Concept and low-latency services strategy}
The main sources of latency in Time Division Multiplexing PON (TDM-PON) \cite{ref_latency1}\cite{ref_latency2}\cite{ref_latency3}, such as GE-PON are: 
(i) Upstream contention between ONUs in the shared medium 
(ii) The efficiency of Dynamic Bandwidth Allocation (DBA)\cite{ref_latency4};
(iii) The ONU discovery and ranging process, where the OLT inserts repeatability a 250–450\,µs quiet windows to the ONUs (for 20–40\,km links).

In the proposed scheme, we consider each PN as a virtual GE-PON, having its own DBA,  but sharing the result from the ONU discovery and ranging process information. 

In contrast to S-VLANs, the virtual GE-PONs have the advantage of not being impacted from a DBA's latency point of view by other virtual GE-PONs. 
Using this concept, a strategy to enable low-latency services is to segment
the data traffic flow of the real GE-PON into a High-Latency Virtual GE-PON (HLV-GE-PON) and a Low-Latency Virtual GE-PON (LLV-GE-PON).

For instance, by directed the data traffic needed for the inevitable "quiet time" to the HLV-GE-PON, this architecture allows services continuity for the ONUs part of other virtual GE-PON during this process. This virtually eliminated the quiet windows for those ONU. 
Additionally,  HLV-GE-PON could be used for other 
non-time-critical systems, 
such as IoT network infrastructure, which have no latency-sensitive machine-to-machine communication.  

Regarding the setting of LLV-GE-PONs,  it should be noted that the reduction number of ONUs in a Virtual GE-PON, inherently result in (i) the reduction other competition of multiple ONUs to receive and transmit data over the Virtual GE-PON and (ii) the reduction of the need for efficient DBA in the Virtual GE-PON.

\section{Conclusion}

A novel low-latency PON network architecture has been proposed,
using OCDMA-based multiple private networks.
In order to address the requirement of some latency-sensitive applications, we define on each private network a Virtual GE-PONs on which the data traffic of the real GE-PON has been redirected based on the services. A strategy has been proposed to ensure the continuity of low-latency services with the setting of dedicated Virtual GE-PONs for low-latency and high-latency. As a result, the proposed architecture is expected to virtually eliminate the inevitable 250–450\,µs\,"quiet time" of ONUs over TDM-PON during the ONU discovery and ranging process.

\begin{credits}

\end{credits}
%
%
%

\begin{thebibliography}{8}

\bibitem{ref_article1}
H. S. Chung, "High Speed and Low Latency PON for 5G Networks," \textit{2019 IEEE Photonics Society Summer Topical Meeting Series (SUM)}, Ft. Lauderdale, FL, USA, 2019, pp. 1-2, \doi:{10.1109/PHOSST.2019.8794896}.

\bibitem{ref_article3}
C. She et al., "Guest Editorial xURLLC in 6G: Next Generation Ultra-Reliable and Low-Latency Communications," in IEEE Journal on Selected Areas in Communications, vol. 41, no. 7, pp. 1963-1968, July 2023, \doi{ 10.1109/JSAC.2023.3282543}.

\bibitem{ref_article4}
C. Bai, P. Dallasega, G. Orzes, and J.Sarkis, “Industry 4.0 technologies assessment: A sustainability perspective,” Int. J. Prod. Econ., vol. 229, p. 107776, 2020, doi: 10.1016/j.ijpe.2020.107776.
  

\bibitem{ref_article2}
Y. Weng, J. Jin, D. Zhang, R. Li, D. Shu and Y. Zuo, "Model and Analysis of the Maximum Upstream Latency of a Deterministic Industrial PON System Applying Frame-Based Dense Burst Allocation Method," in Journal of Lightwave Technology, vol. 42, no. 9, pp. 3210-3220, 1 May1, 2024, \doi{ 10.1109/JLT.2024.3355440}.

\bibitem{ref_Gh_1}
M. Gharaei, S. Cordette, C. Lepers, I. Fsaifes, and P. Gallion,"Multiple Optical Private Networks Over EPON Using Optical CDMA Technique," in \textit{National Fiber Optic Engineers Conference}, OSA Technical Digest (CD) (Optica Publishing Group, 2010), paper JThA34.Author, A.-B.: Contribution title. In: 9th International Proceedings
on Proceedings, pp. 1--2. Publisher, Location (2010) \doi{10.1364/NFOEC.2010.JThA34}

\bibitem{ref_Gh_2}
M. Gharaei, S. Cordette, P. Gallion, C. Lepers and I. Fsaifes, "Enabling internetworking among ONUs in EPON using OCDMA technique," \textit{2009 3rd International Conference on Signals, Circuits and Systems (SCS)}, Medenine, Tunisia, 2009, pp. 1-4, \doi{10.1109/ICSCS.2009.5412582}.

\bibitem{ref_Ihs}
I. Fsaifes, C. Lepers, M. Lourdiane, P. Gallion, V. Beugin and P. Guignard, “Source Coherence Impairments in a Direct Detection DSOCDMA system,” Applied Optics, vol. 46, no. 4, pp. 456-462, Feb. 2007.

\bibitem{ref_ware1}
C. Ware \textit{et al}.: Spectral Slicing of a Supercontinuum Source for WDM/DS-OCDMA Application. \textit{2008 10th Anniversary International Conference on Transparent Optical Networks}, Athens, Greece, 2008, pp. 158-161, \doi{ 10.1109/ICTON.2008.4598758}.

\bibitem{ref_ware2}
C. Ware et al., "Optical CDMA enhanced by nonlinear optics," 2010 12th International Conference on Transparent Optical Networks, Munich, Germany, 2010, pp. 1-4, \doi{10.1109/ICTON.2010.5548965}

\bibitem{ref_NLT_Ihs_1}
I. Fsaifes \textit{et al}., "Nonlinear Pulse Reshaping With Highly Birefringent Photonic Crystal Fiber for OCDMA Receivers," in \textit{IEEE Photonics Technology Letters}, vol. 22, no. 18, pp. 1367-1369, Sept.15, 2010, \doi{ 10.1109/LPT.2010.2057415}.

\bibitem{ref_NLT_1}
X. Wang, N. Wada, K. Kitayama, “Performance degradation in coherent OCDMA due to receiver’s bandwidth limit and improvement by using optical thresholding,” in proc. LEOS, pp. 638- 639, Oct. 2005.

\bibitem{ref_NLT_2}
C. Huang \textit{et al}., "Programmable Silicon Photonic Optical Thresholder," in \textit{IEEE Photonics Technology Letters}, vol. 31, no. 22, pp. 1834-1837, 15 Nov.15, 2019, \doi{10.1109/LPT.2019.2948903}.

\bibitem{ref_NLT_3}
A. R. Forouzan, M. Nasiri-Kenari, J. A. Salehi and N. Rezaee, "Chip-level detector with optimum comparator threshold and correlation receiver with an electrical hard-limiter for optical cdma," \textit{IEEE International Symposium on Information Theory, 2003. Proceedings.}, Yokohama, Japan, 2003, pp. 326-326, \doi{10.1109/ISIT.2003.1228341}.

\bibitem{ref_latency1}
W. L. Zhang and L. Q. Yuan: Higher speed passive optical networks for low latency services. ZTE Communications, vol. 19, no. 2, pp. 61–66, Jun. 2021. \doi{10.12142/ZTECOM.202102008}

\bibitem{ref_latency2}
K. Kim, K. Doo and H. Chung, "Demonstration of Low Latency 25G TDM-PON with Flexible Multizone-based ONU Activation for Time Critical Services," 2022 European Conference on Optical Communication (ECOC), Basel, Switzerland, 2022, pp. 1-3. 

\bibitem{ref_latency3}
S. Hatta, N. Tanaka, and T. Sakamoto, “Low Latency Dynamic Bandwidth Allocation Method with High Bandwidth Efficiency for TDM-PON,” \textit{NTT Tech. Rev.}, vol. 15, no. 4, pp. 50–56, 2017, \doi{10.53829/ntr201704ra1}.

\bibitem{ref_latency4}
P. S. C. Chitra, A. Samanta, A. K. Dutta and A. Adhya, "Efficient Low-Latency and Traffic Aware Dynamic Bandwidth Allocation for XG-PON," 2024 IEEE International Conference on Advanced Networks and Telecommunications Systems (ANTS), Guwahati, India, 2024, pp. 1-6, \doi{10.1109/ANTS63515.2024.10898466}.



\end{thebibliography}
%

\end{document}